\documentclass[10pt,conference]{IEEEtran}
\IEEEoverridecommandlockouts
\usepackage{amsmath,amssymb,amsfonts}
\usepackage{algorithmic}
\usepackage{graphicx}
\usepackage{xcolor}
\usepackage{amsmath,lipsum}
\usepackage[ruled]{algorithm2e}
\usepackage{float}
\usepackage{subfigure}
\usepackage{multirow}
\usepackage{multicol}
\usepackage{cite}
\usepackage{stfloats}
\usepackage{booktabs}
\usepackage{caption}
\usepackage{textcomp}
\usepackage{arydshln}
\usepackage{makecell}
\usepackage{bm}
\usepackage[explicit]{titlesec}

\DeclareMathOperator*{\argmin}{arg\,min}

\def\BibTeX{{\rm B\kern-.05em{\sc i\kern-.025em b}\kern-.08em
    T\kern-.1667em\lower.7ex\hbox{E}\kern-.125emX}}
\begin{document}

\title{Personalized Digital Semantic Communication for Image Transmission with Vision-Language Models}

\author{
\IEEEauthorblockN{Nan Li, Li Zhou, Haijun Wang, Jun Xiong, Haitao Zhao, Jibo Wei}
\IEEEauthorblockA{
College of Electronic Science, National University of Defense Technology, China\\
Email: \{li.nan, zhouli2035, haijunwang14, xj8765, haitaozhao, wjbhw\}@nudt.edu.cn}
}

\maketitle

\begin{abstract}
Semantic communication (SC) enables bandwidth-efficient wireless image transmission, but most existing SC schemes are user-agnostic and ignore receiver-dependent semantics. To address this issue, we propose a personalized digital semantic communication (PDSC) framework that integrates a vision-language model (VLM)-based semantic encoder with a latent diffusion model (LDM)-based semantic decoder. Specifically, the semantic encoder extracts source-aware personalized semantic tokens from both the source image and the receiver's historical interactions. These tokens are vector-quantized into discrete semantic indices and further encoded into a compact fixed-length bitstream, enabling compatibility with digital transmission. At the receiver, the semantic decoder reconstructs a personalized image conditioned on the recovered semantic tokens. Furthermore, we formulate a capacity-constrained personalized semantic rate-distortion problem and introduce a semantic distortion metric that jointly characterizes source-semantic fidelity and user-preference alignment. Experiments show that PDSC achieves superior source-semantic consistency and personalization over state-of-the-art SC baselines, including CDDM and MoS, under bandwidth-limited wireless transmission.
\end{abstract}


\section{Introduction}

The rapid growth of visual-intensive applications, such as augmented reality and personalized content delivery, has imposed stringent requirements on bandwidth-efficient and low-latency wireless image transmission. Conventional communication systems, which aim to reliably transmit bitstreams under Shannon's technical framework, usually preserve pixel-level fidelity without explicitly exploiting the semantic relevance of visual content. To reduce redundant and task-irrelevant information, \textit{semantic communication} (SC) has emerged as a promising paradigm that transmits the meaning or task-relevant information rather than the raw data~\cite{luo2022semantic}.

Despite recent advances in image semantic communication systems~\cite{jiang2022semantic}, most existing frameworks remain user-agnostic. They typically adopt a rigid \emph{one-size-fits-all} strategy, where semantic importance is assumed to be universal and independent of the receiver. Consequently, a single semantic representation is learned and optimized for generic reconstruction or downstream tasks. In practical personalized services, however, semantic relevance is inherently receiver-dependent: the information that is useful, interesting, or actionable may vary significantly across users. For example, the same movie poster may attract different attention from users who prefer action movies, romantic movies, or animation. This observation motivates semantic communication systems that are not only content-aware but also receiver-aware~\cite{gunduz2023beyond}.

Recent advances in generative artificial intelligence, particularly diffusion models~\cite{podell2024sdxl}, offer new opportunities for receiver-side semantic reconstruction. Diffusion-based generative SC frameworks~\cite{wu2025cddm,ni2025mixture,li2026goal} have shown that compact semantic features transmitted over noisy channels can be transformed into high-quality images by leveraging powerful pretrained generative priors. For example, CDDM~\cite{wu2025cddm} employs a channel denoising diffusion model at the receiver to remove channel-induced perturbations from received semantic features. In \cite{ni2025mixture}, MoS adopts a mixture-of-semantics strategy to adaptively allocate bandwidth to regions of interest and non-interest, thereby improving transmission efficiency. However, these methods mainly focus on universal reconstruction quality and do not consider the receiver's personal preference. As a result, the reconstructed images may preserve general source semantics but fail to reflect user-specific preferences.

To fill these gaps, 
we propose a vision-language model (VLM)-enabled \emph{Personalized Digital Semantic Communication} (PDSC) framework, with the goal of reducing transmission overhead while reconstructing images that are both semantically faithful to the source and aligned with the receiver's visual preference. The main contributions are:
\begin{itemize}
    \item We propose a VLM-based personalized semantic encoder that extracts preference, visual, and textual semantic tokens to capture receiver-dependent preference, source structure, and high-level semantic guidance, respectively.
    
    \item We develop a vector-quantized interface to convert personalized semantic tokens into a fixed-length bitstream, and an LDM-based decoder to reconstruct personalized images from the received semantics.
    
    \item We formulate a capacity-constrained personalized semantic rate-distortion problem and introduce a semantic distortion metric jointly measuring source-semantic fidelity and user-preference alignment.  Experiments show that PDSC outperforms state-of-the-art SC baselines, CDDM and MoS, in semantic consistency and personalization under bandwidth-limited wireless transmission.

\end{itemize}

\begin{figure*}[t]
    \centering
    \includegraphics[width=0.95\linewidth]{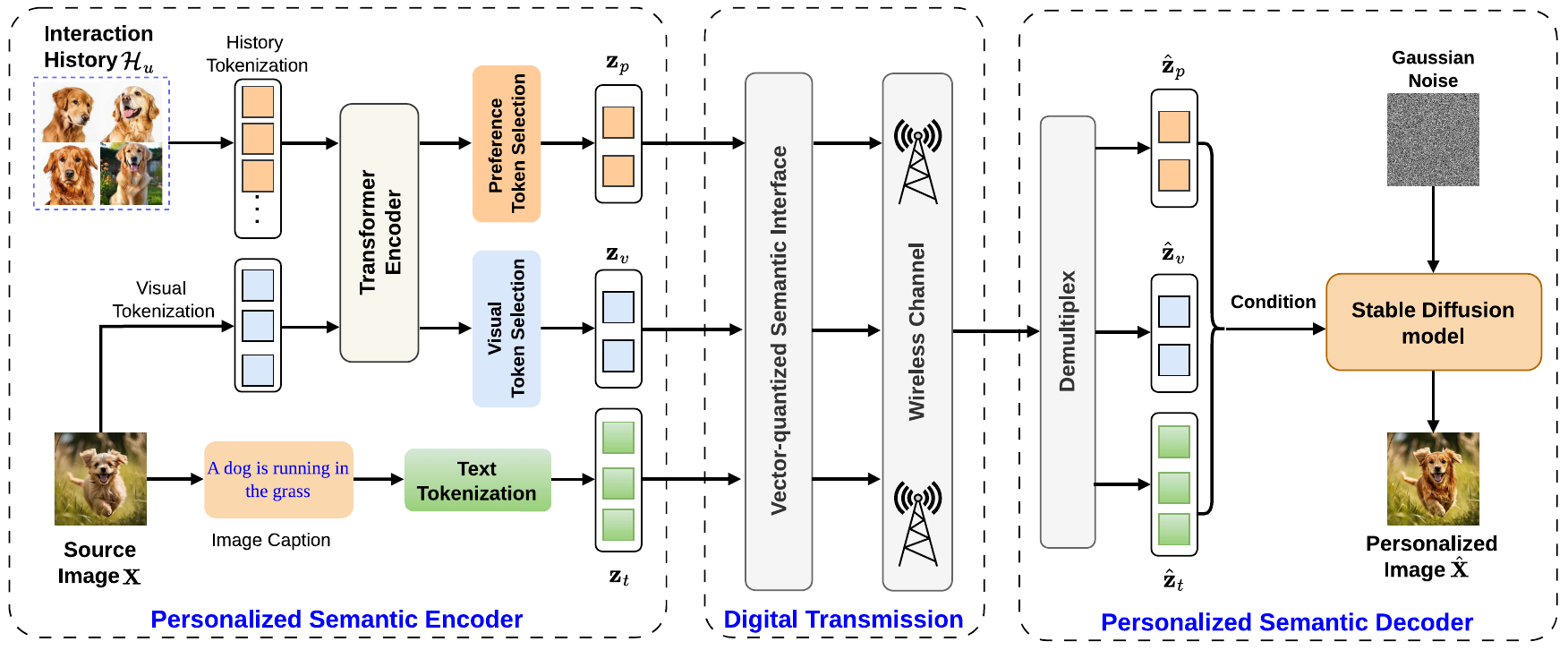}
    \caption{The proposed personalized digital semantic communication (PDSC) framework for wireless image transmission.}
    \label{fig:system}
\end{figure*}

\section{System Model and Problem Formulation}
We consider an end-to-end wireless image transmission task between a transmitter and a receiver, as illustrated in Fig.~\ref{fig:system}. Let $\mathbf{X} \in \mathbb{R}^{H \times W \times C}$ denote the source image to be transmitted, where $H$, $W$, and $C$ denote the height, width, and number of channels, respectively. The receiver is characterized by a historical interaction set $\mathcal{H}_u = \{\mathbf{x}_1,\ldots,\mathbf{x}_N\}$ that reflects its long-term visual preferences. The communication goal is to reconstruct an image $\hat{\mathbf{X}}$ that preserves the high-level semantics of $\mathbf{X}$ while being aligned with the personalized visual preference implied by $\mathcal{H}_u$. To this end, we propose a personalized digital semantic communication (PDSC) framework consisting of a vision-language model (VLM)-based semantic encoder at the transmitter, a vector quantization and fixed-length coding module for digital transmission, and a latent diffusion model (LDM)-based semantic decoder at the receiver.

\subsection{Personalized Semantic Codec}

The semantic encoder extracts compact personalized semantic tokens by conditioning on both the source image $\mathbf{X}$ and the user history $\mathcal{H}_u$. This process is modeled as 
\begin{equation}
  (\mathbf{z}_p, \mathbf{z}_v, \mathbf{z}_t) = \mathcal{E}_{\phi}(\mathbf{X}, \mathcal{H}_u),
  \label{eq:encoder}
\end{equation}
where $\mathcal{E}_{\phi}(\cdot)$ denotes the personalized semantic encoder with trainable parameters $\phi$. Here, $\mathbf{z}_p$ denotes the preference token that captures source-relevant receiver preference, $\mathbf{z}_v$ denotes visual-semantic tokens that preserve the structural and visual information of the source image, and $\mathbf{z}_t$ denotes text-semantic tokens that provide explicit high-level semantic guidance.

The three types of semantic tokens are projected into a common latent space and concatenated as
\begin{equation}
    \mathbf{Z} = \mathrm{Concat}\left[\mathbf{z}_p,\mathbf{z}_v,\mathbf{z}_t\right]
    \in \mathbb{R}^{L	\times d},
    \label{eq:concat}
\end{equation}
where $L$ is the total number of transmitted semantic tokens and $d$ is the token dimension. 

To enable digital transmission, the continuous semantic tokens are converted into discrete semantic indices through vector quantization. Let $\mathcal{C}=\{\mathbf{c}_1,\ldots,\mathbf{c}_K\}\subset\mathbb{R}^{d}$ denote a learnable semantic codebook with $K$ codewords. Each semantic token $\mathbf{z}_j$ is mapped to its nearest codeword by
\begin{equation}
    k_j = \argmin_{k\in\{1,\ldots,K\}} \|\mathbf{z}_j-\mathbf{c}_k\|_2, \quad \tilde{\mathbf{z}}_j = \mathbf{c}_{k_j}.
    \label{eq:vq}
\end{equation}
The resulting semantic index sequence is denoted by $\mathbf{k}=\{k_1,\ldots,k_L\}$. With fixed-length binary coding, the semantic source rate is given by
\begin{equation}
    R_s = L\lceil\log_2 K\rceil,
    \label{eq:semantic_rate}
\end{equation}
where $L$ controls the number of transmitted semantic tokens and $K$ controls the quantization resolution of each token.

We consider a complex channel with $N_{\mathrm{ch}}$ channel uses per image. Under channel signal-to-noise ratio $\mathrm{SNR}$, the maximum reliably supported payload per image is
\begin{equation}
    C_{\mathrm{ch}} = N_{\mathrm{ch}}\log_2(1+\mathrm{SNR}).
    \label{eq:channel_budget}
\end{equation}

After channel decoding, the receiver obtains $\hat{\mathbf{k}}=\{\hat{k}_1,\ldots,\hat{k}_L\}$ and retrieves the quantized tokens by codebook lookup, i.e.,
$\hat{\mathbf{z}}_j=\mathbf{c}_{\hat{k}_j}$ for $j=1,\ldots,L$.
The recovered tokens are split into $\hat{\mathbf{z}}_p$, $\hat{\mathbf{z}}_v$, and $\hat{\mathbf{z}}_t$, and decoded to reconstruct the personalized image as
\begin{equation}
    \hat{\mathbf{X}}=
    \mathcal{D}_{\theta}(\hat{\mathbf{z}}_p,\hat{\mathbf{z}}_v,\hat{\mathbf{z}}_t),
    \label{eq:decoder}
\end{equation}
where $\mathcal{D}_{\theta}(\cdot)$ denotes the diffusion-based semantic decoder.

\subsection{Personalized Semantic Distortion}

To evaluate the reconstruction quality from both the source and receiver perspectives, we define a personalized semantic distortion metric that jointly accounts for source-semantic fidelity and user-preference alignment:
\begin{equation}
    D(\mathbf{X},\hat{\mathbf{X}},\mathcal{H}_u)
    =
    \alpha L_{\mathrm{sem}}(\mathbf{X},\hat{\mathbf{X}})
    +
    (1-\alpha)L_{\mathrm{pref}}(\hat{\mathbf{X}},\mathcal{H}_u),
    \label{eq:distortion}
\end{equation}
where $\alpha\in[0,1]$ controls the tradeoff between preserving source semantics and matching user preference. The source-semantic fidelity term is defined as the cosine distance in a pretrained vision-language embedding space:
\begin{equation}
    L_{\mathrm{sem}}(\mathbf{X},\hat{\mathbf{X}})
    =
    1 -
    \frac{
    \Phi_I(\mathbf{X})^{\mathsf{T}}\Phi_I(\hat{\mathbf{X}})
    }{
    \|\Phi_I(\mathbf{X})\|_2
    \|\Phi_I(\hat{\mathbf{X}})\|_2
    },
    \label{eq:lsem}
\end{equation}
where $\Phi_I(\cdot)$ denotes the image encoder of a pretrained VLM.

The preference alignment term measures the consistency between the reconstructed image and the receiver's source-relevant historical preferences. To avoid treating all historical images equally, we assign each historical image a source-conditioned relevance weight:
\begin{equation}
    w_i =
    \frac{
    \exp\left(
    \mathrm{sim}_{I,I}(\mathbf{X},\mathbf{x}_i)/\tau
    \right)
    }{
    \sum_{j=1}^{N}
    \exp\left(
    \mathrm{sim}_{I,I}(\mathbf{X},\mathbf{x}_j)/\tau
    \right)
    },
    \label{eq:weight}
\end{equation}
where $\tau>0$ is a temperature parameter and
\begin{equation}
    \mathrm{sim}_{I,I}(\mathbf{a},\mathbf{b})
    =
    \frac{
    \Phi_I(\mathbf{a})^{\mathsf{T}}\Phi_I(\mathbf{b})
    }{
    \|\Phi_I(\mathbf{a})\|_2
    \|\Phi_I(\mathbf{b})\|_2
    }.
\end{equation}
The preference alignment loss is then given by
\begin{equation}
    L_{\mathrm{pref}}(\hat{\mathbf{X}},\mathcal{H}_u)
    =
    1 -
    \sum_{i=1}^{N}
    w_i\,
    \mathrm{sim}_{I,I}(\hat{\mathbf{X}},\mathbf{x}_i).
    \label{eq:lpref}
\end{equation}
Unlike conventional pixel-wise distortion metrics, the proposed distortion explicitly captures receiver-dependent semantic relevance by emphasizing historical preferences that are related to the current source image.

\subsection{Personalized Semantic Rate-Distortion Problem}

Under the fixed-length digital semantic interface, the system aims to minimize the expected personalized semantic distortion subject to the channel-dictated rate constraint. For a given semantic configuration $(L,K)$, the personalized semantic rate-distortion optimization problem is formulated as
\begin{equation}
\begin{aligned}
    \min_{\Theta} \quad
    & \mathbb{E}_{\mathbf{X},\mathcal{H}_u}
    \left[
    D(\mathbf{X},\hat{\mathbf{X}},\mathcal{H}_u)
    \right] \\
    \mathrm{s.t.} \quad
    & R_s \le C_{\mathrm{ch}}.
\end{aligned}
    \label{eq:opt}
\end{equation}
where $\Theta = \{\phi,\theta,\mathcal{C}\}$
denotes the set of trainable parameters. The pair $(L,K)$ determines the semantic operating point, where increasing $L$ transmits more semantic tokens, while increasing $K$ reduces quantization distortion for each token. 

\section{Framework of the Proposed PDSC}

This section presents the implementation of PDSC, including the personalized semantic encoder, vector-quantized digital interface, diffusion-based decoder, and training strategy.

\subsection{Personalized Semantic Encoder}
From a communication-theoretic perspective, the personalized semantic encoder acts as a semantic source encoder to extract a compact representation sufficient to minimize the personalized semantic distortion. The receiver’s historical interactions serve as semantic side information, allowing the encoder to suppress receiver-irrelevant visual information while retaining source attributes valuable to the intended user.

First, a frozen visual tokenizer extracts visual token sequences from the source image and historical images:
\begin{equation}
    \mathbf{E}_0 = \mathrm{VT}(\mathbf{X}) \in \mathbb{R}^{L_0 \times d_{\mathrm{vis}}},
\end{equation}
\begin{equation}
    \mathbf{E}_i = \mathrm{VT}(\mathbf{x}_i) \in \mathbb{R}^{L_i \times d_{\mathrm{vis}}},
    \quad i=1,\ldots,N,
\end{equation}
The historical visual tokens are then concatenated as
\begin{equation}
    \mathbf{E}_{\mathcal{H}} =
    \mathrm{Concat}\left[\mathbf{E}_1,\ldots,\mathbf{E}_N\right]
    \in \mathbb{R}^{L_{\mathcal{H}}\times d_{\mathrm{vis}}}.
\end{equation}

Directly using all historical tokens is inefficient and may introduce preference information irrelevant to the current source image. Therefore, we employ a lightweight context filtering module to retain source-relevant historical tokens and preference-relevant source tokens. The source and historical tokens are first projected into a common hidden space and jointly processed by a Transformer encoder:
\begin{equation}
    (\mathbf{Z}_0,\mathbf{Z}_{\mathcal{H}})
    =
    \mathrm{Trans}
    \left(
    \mathrm{Proj}_0(\mathbf{E}_0),
    \mathrm{Proj}_{\mathcal{H}}(\mathbf{E}_{\mathcal{H}})
    \right),
\end{equation}
where $\mathbf{Z}_0\in\mathbb{R}^{L_0	\times d}$ and
$\mathbf{Z}_{\mathcal{H}}\in\mathbb{R}^{L_{\mathcal{H}}\times d}$ are context-enhanced source and history representations, respectively.

The relevance score of each historical token is computed according to its maximum similarity to the source tokens:
\begin{equation}
    s_j^{\mathcal{H}}
    =
    \max_{\ell\in\{1,\ldots,L_0\}}
    \cos\left(\mathbf{Z}_{\mathcal{H}}[j],\mathbf{Z}_{0}[\ell]\right),
    \quad j=1,\ldots,L_{\mathcal{H}}.
\end{equation}
Similarly, the relevance score of each source token with respect to the user history is defined as
\begin{equation}
    s_\ell^{0}
    =
    \max_{j\in\{1,\ldots,L_{\mathcal{H}}\}}
    \cos\left(\mathbf{Z}_{0}[\ell],\mathbf{Z}_{\mathcal{H}}[j]\right),
    \quad \ell=1,\ldots,L_0.
\end{equation}
Based on these scores, top-ranked tokens are retained to form binary masks $\mathbf{m}_{h}$ and $\mathbf{m}_{s}$. During training, the Gumbel-Softmax reparameterization is adopted to make the masking operation differentiable. During inference, hard top-$M$ selection is used. The filtered token sets are denoted by
\begin{equation}
    \tilde{\mathbf{E}}_{\mathcal{H}}
    =
    \mathrm{Select}(\mathbf{E}_{\mathcal{H}},\mathbf{m}_{h}),
    \quad
    \tilde{\mathbf{E}}_{0}
    =
    \mathrm{Select}(\mathbf{E}_{0},\mathbf{m}_{s}).
\end{equation}

The preference token is extracted by querying the filtered historical tokens with the filtered source tokens through a cross-attention module:
\begin{equation}
    \mathbf{A}_p =
    \mathrm{Softmax}
    \left(
    \frac{
    (\tilde{\mathbf{E}}_0\mathbf{W}^{Q})
    (\tilde{\mathbf{E}}_{\mathcal{H}}\mathbf{W}^{K})^{\mathsf{T}}
    }{
    \sqrt{d_k}
    }
    \right)
    \tilde{\mathbf{E}}_{\mathcal{H}}\mathbf{W}^{V},
\end{equation}
where $\mathbf{W}^{Q}$, $\mathbf{W}^{K}$, and $\mathbf{W}^{V}$ are learnable projection matrices. The preference token is obtained by average pooling and nonlinear projection:
\begin{equation}
    \mathbf{z}_p =
    \mathrm{MLP}_p
    \left(
    \mathrm{AvgPool}(\mathbf{A}_p)
    \right)
    \in \mathbb{R}^{L_p	\times d}.
\end{equation}

The visual-semantic tokens are extracted from the filtered source tokens to preserve source structure and visual layout. Instead of directly transmitting dense visual features, we use a set of learnable visual queries $\mathbf{Q}_v\in\mathbb{R}^{L_v	\times d}$ to aggregate compact source information:
\begin{equation}
    \mathbf{z}_v =
    \mathrm{CrossAttn}
    \left(
    \mathbf{Q}_v,
    \mathrm{Proj}_{v}(\tilde{\mathbf{E}}_0),
    \mathrm{Proj}_{v}(\tilde{\mathbf{E}}_0)
    \right)
    \in \mathbb{R}^{L_v	\times d}.
\end{equation}

\addtolength{\topmargin}{0.01in}

To provide explicit high-level semantic guidance, a captioning model generates a textual description of the source image. The generated text is then encoded and projected into the common semantic space:
\begin{equation}
    \mathbf{z}_t =
    \mathrm{Proj}_{t}
    \left(
    \mathrm{TextEnc}
    \left(
    \mathrm{BLIP}(\mathbf{X})
    \right)
    \right)
    \in \mathbb{R}^{L_t	\times d},
\end{equation}
where $\mathrm{BLIP}(\cdot)$ denotes the image captioning module and $\mathrm{TextEnc}(\cdot)$ denotes the text encoder.

\subsection{Vector-Quantized Digital Semantic Interface}

The continuous tokens $\mathbf{Z}$ are quantized into indices $\mathbf{k}$ using the codebook $\mathcal{C}$. Since nearest-neighbor quantization is non-differentiable, we use the straight-through estimator~\cite{oord2017vqvae}:
\begin{equation}
    \tilde{\mathbf{z}}_j
    =
    \mathbf{z}_j
    +
    \mathrm{sg}
    \left[
    \mathbf{c}_{k_j}-\mathbf{z}_j
    \right],
\end{equation}
where $\mathrm{sg}[\cdot]$ denotes stop-gradient. In the forward pass, $\tilde{\mathbf{z}}_j$ equals the selected codeword $\mathbf{c}_{k_j}$, while in the backward pass, gradients are copied from the quantized token to the encoder output. The VQ loss is expressed as
\begin{equation}
\begin{aligned}
    \mathcal{L}_{\mathrm{vq}}
    =
    \frac{1}{L}
    \sum_{j=1}^{L}
    \left\|
    \mathrm{sg}[\mathbf{z}_j]
    -
    \mathbf{c}_{k_j}
    \right\|_2^2
    +
    \beta_{\mathrm{com}}
    \frac{1}{L}
    \sum_{j=1}^{L}
    \left\|
    \mathbf{z}_j
    -
    \mathrm{sg}[\mathbf{c}_{k_j}]
    \right\|_2^2,
\end{aligned}
\end{equation}
where the first term updates the codebook, while the second encourages the outputs to stay close to the selected codewords; $\beta_{\mathrm{com}}$ controls the strength of the commitment term.

To avoid codebook collapse and promote balanced codeword usage, we further introduce an entropy-based regularizer:
\begin{equation}
    \mathcal{L}_{\mathrm{u}}
    =
    \sum_{k=1}^{K}
    \hat{p}_k
    \log(\hat{p}_k+\epsilon),
\end{equation}
where $\hat{p}_k$ is the empirical usage frequency of the $k$-th codeword in a mini-batch and $\epsilon$ is a small constant for numerical stability. Minimizing $\mathcal{L}_{\mathrm{u}}$ encourages a high-entropy codeword distribution and alleviates codebook collapse.

At the receiver, after channel decoding, the recovered index sequence $\hat{\mathbf{k}}=\{\hat{k}_1,\ldots,\hat{k}_L\}$ is mapped back to quantized tokens $\hat{\mathbf{z}}_j=\mathbf{c}_{\hat{k}_j}$ through codebook lookup. The recovered token sequence is then split into $\hat{\mathbf{z}}_p$, $\hat{\mathbf{z}}_v$, and $\hat{\mathbf{z}}_t$ according to the predefined token layout.

\subsection{Diffusion-Based Semantic Decoder}

The semantic decoder reconstructs images from the recovered tokens using an LDM prior. Let $\mathbf{x}_0$ be the latent representation of the target image. The forward diffusion process is
\begin{equation}
    \mathbf{x}_t
    =
    \sqrt{\bar{\alpha}_t}\mathbf{x}_0
    +
    \sqrt{1-\bar{\alpha}_t}\boldsymbol{\epsilon},
    \quad
    \boldsymbol{\epsilon}\sim\mathcal{N}(\mathbf{0},\mathbf{I}),
\end{equation}
where $t$ is the diffusion step and $\bar{\alpha}_t$ denotes the cumulative noise schedule. The denoising network predicts the injected noise conditioned on the recovered semantic tokens:
\begin{equation}
    \hat{\boldsymbol{\epsilon}}
    =
    \epsilon_{\theta}
    \left(
    \mathbf{x}_t,
    t,
    \hat{\mathbf{z}}_p,
    \hat{\mathbf{z}}_v,
    \hat{\mathbf{z}}_t
    \right),
\end{equation}
where $\epsilon_{\theta}(\cdot)$ denotes the conditional denoising network.

Following the conditioning design of diffusion models~\cite{rombach2022high}, $\hat{\mathbf{z}}_p$ is injected through cross-attention to guide personalized attributes, $\hat{\mathbf{z}}_v$ modulates U-Net features via feature-wise linear modulation to preserve structure, and $\hat{\mathbf{z}}_t$ provides text-level semantic guidance. This multi-path conditioning design enables the decoder to jointly balance personalization, structural preservation, and semantic consistency.

During training, the diffusion denoising loss is given by
\begin{equation}
    \mathcal{L}_{\mathrm{diff}}
    =
    \mathbb{E}_{t,\boldsymbol{\epsilon}}
    \left[
    \left\|
    \boldsymbol{\epsilon}
    -
    \epsilon_{\theta}
    \left(
    \mathbf{x}_t,
    t,
    \tilde{\mathbf{z}}_p,
    \tilde{\mathbf{z}}_v,
    \tilde{\mathbf{z}}_t
    \right)
    \right\|_2^2
    \right],
\end{equation}
where $\tilde{\mathbf{z}}_p$, $\tilde{\mathbf{z}}_v$, and $\tilde{\mathbf{z}}_t$ are the quantized semantic tokens produced by the straight-through estimator.

At inference, the personalized image is generated by deterministic or stochastic diffusion sampling:
\begin{equation}
    \hat{\mathbf{X}}
    =
    \mathcal{D}_{\theta}
    \left(
    \hat{\mathbf{z}}_p,
    \hat{\mathbf{z}}_v,
    \hat{\mathbf{z}}_t
    \right),
\end{equation}
where $\mathcal{D}_{\theta}(\cdot)$ denotes the full LDM-based semantic decoder.

\begin{algorithm}[t]
\caption{Training and Inference of PDSC}
\label{alg:pdsc}
\begin{algorithmic}[1]
\REQUIRE Training data $\mathcal{D}$, token number $L$, codebook size $K$, codebook $\mathcal{C}$, loss weights $\lambda_{\mathrm{vq}}$, $\lambda_{\mathrm{u}}$, $\lambda_{\mathrm{s}}$
\ENSURE Trained semantic encoder $\mathcal{E}_{\phi}$, codebook $\mathcal{C}$, and semantic decoder $\mathcal{D}_{\theta}$

\tcp{\textbf{\textit{Semantic token alignment}}}

\STATE Freeze the pretrained visual tokenizer, VLM components, and diffusion backbone.
\REPEAT
    \STATE Sample a mini-batch $(\mathbf{X},\mathcal{H}_u)$ from $\mathcal{D}$.
    \STATE Extract semantic tokens $(\mathbf{z}_p,\mathbf{z}_v,\mathbf{z}_t)=\mathcal{E}_{\phi}(\mathbf{X},\mathcal{H}_u)$.
    \STATE Quantize semantic tokens into indices $\mathbf{k}$ and tokens $(\tilde{\mathbf{z}}_p,\tilde{\mathbf{z}}_v,\tilde{\mathbf{z}}_t)$.
    \STATE Compute $\mathcal{L}_{\mathrm{I}}$.
    \STATE Update $\mathcal{E}_{\phi}$, projection layers, and $\mathcal{C}$.
\UNTIL{convergence}

\tcp{\textbf{\textit{Personalization adaptation}}}

\STATE Unfreeze the condition-injection modules of $\mathcal{D}_{\theta}$.
\REPEAT
    \STATE Sample a mini-batch $(\mathbf{X},\mathcal{H}_u)$ from $\mathcal{D}$.
    \STATE Extract and quantize semantic tokens as in Stage I.
    \STATE Generate $\hat{\mathbf{X}}=\mathcal{D}_{\theta}(\tilde{\mathbf{z}}_p,\tilde{\mathbf{z}}_v,\tilde{\mathbf{z}}_t)$ or its efficient approximation.
    \STATE Compute $\mathcal{L}_{\mathrm{II}}$.
    \STATE Update $\mathcal{E}_{\phi}$, $\mathcal{C}$, and the condition-injection modules.
\UNTIL{convergence}

\tcp{\textbf{\textit{Inference}}}

\STATE Extract and quantize semantic tokens into indices $\mathbf{k}$.
\STATE Encode $\mathbf{k}$ into a fixed-length bitstream and transmit it with channel coding and digital modulation.
\STATE Recover $\hat{\mathbf{k}}$ after channel decoding and obtain the corresponding semantic tokens.
\STATE Reconstruct $\hat{\mathbf{X}}=\mathcal{D}_{\theta}(\hat{\mathbf{z}}_p,\hat{\mathbf{z}}_v,\hat{\mathbf{z}}_t)$.
\end{algorithmic}
\end{algorithm}

\subsection{Training and Inference}

To improve convergence and training stability, we adopt a two-stage training strategy. First, we freeze the pretrained visual tokenizer, VLM components, and diffusion backbone, and train only the semantic encoder, projection layers, and codebook using the loss:
\begin{equation}
    \mathcal{L}_{\mathrm{I}}
    =
    \mathcal{L}_{\mathrm{diff}}
    +
    \lambda_{\mathrm{vq}}\mathcal{L}_{\mathrm{vq}}
    +
    \lambda_{\mathrm{u}}\mathcal{L}_{\mathrm{u}},
\end{equation}
where $\lambda_{\mathrm{vq}}$ and $\lambda_{\mathrm{u}}$ balance the VQ loss and the usage regularizer, respectively. This stage forces the discrete semantic codewords to become compatible with the pretrained diffusion prior.

We then unfreeze the condition-injection modules of the diffusion decoder and jointly fine-tune them with the semantic encoder and codebook. The objective is augmented with a personalized semantic distortion:
\begin{equation}
    \mathcal{L}_{\mathrm{II}}
    =
    \mathcal{L}_{\mathrm{diff}}
    +
    \lambda_{\mathrm{vq}}\mathcal{L}_{\mathrm{vq}}
    +
    \lambda_{\mathrm{u}}\mathcal{L}_{\mathrm{u}}
    +
    \lambda_{\mathrm{s}}
    D(\mathbf{X},\hat{\mathbf{X}},\mathcal{H}_u).
\end{equation}
where the distortion term is computed using the predicted clean latent or a low-step decoded reconstruction to reduce training cost.

During inference, the transmitter extracts and quantizes personalized semantic tokens, encodes the indices into a bitstream, and transmits them with conventional channel coding and modulation. The receiver recovers the indices, retrieves the codewords, and reconstructs the image using the diffusion decoder. Algorithm~\ref{alg:pdsc} summarizes the procedure.

\section{Performance Evaluation}

We evaluate PDSC for wireless image transmission over an AWGN channel. Experiments are conducted using MovieLens-Latest user-rating records \cite{harper2015movielens} with the corresponding movie poster images collected from public metadata. For each receiver, we select $N=5$ positively rated posters to form the historical preference set $\mathcal{H}_u$. The VLM module is implemented based on LaVIT \cite{jin2024lavit}, and the semantic decoder is built upon Stable Diffusion XL~\cite{podell2024sdxl}. Unless otherwise specified, we set $L_p=1$, $L_v=15$, $L_t=5$, and thus $L=21$. The codebook size is $K=1024$. The model is trained with Adam using a learning rate of $10^{-4}$, and the loss weights are $\lambda_{\mathrm{vq}}=1.0$, $\lambda_{\mathrm{u}}=0.1$, and $\lambda_{\mathrm{s}}=1.0$.

\subsection{Baselines and Metric}

We compare PDSC with three representative baselines: 
1) \textbf{BPG+LDPC}, a separation-based scheme where the source image is compressed by BPG and protected by LDPC channel coding; 
2) \textbf{CDDM}\cite{wu2025cddm}, a diffusion-based semantic communication method using a channel denoising diffusion model for reconstruction; and 
3) \textbf{MoS}\cite{ni2025mixture}, a mixture-of-semantics method that transmits compact semantic features for generative reconstruction. For fair comparison, all methods are constrained to the same information bit budget per image before channel coding.

To measure the personalized semantic reconstruction quality, we adopt two metrics: 1) \textbf{CLIP Image Score (CIS)} measures the CLIP image similarity between the reconstructed image and the receiver's historical preference images, reflecting user-preference alignment; 2) \textbf{CLIP Score (CS)} measures semantic consistency between the reconstructed image and the textual descriptions of the source image. Higher CIS and CS indicate better personalization and source-semantic preservation, respectively.

\begin{figure}[t]
    \centering
    \subfigure[CIS versus SNR.]{
    \includegraphics[width=0.4\textwidth]{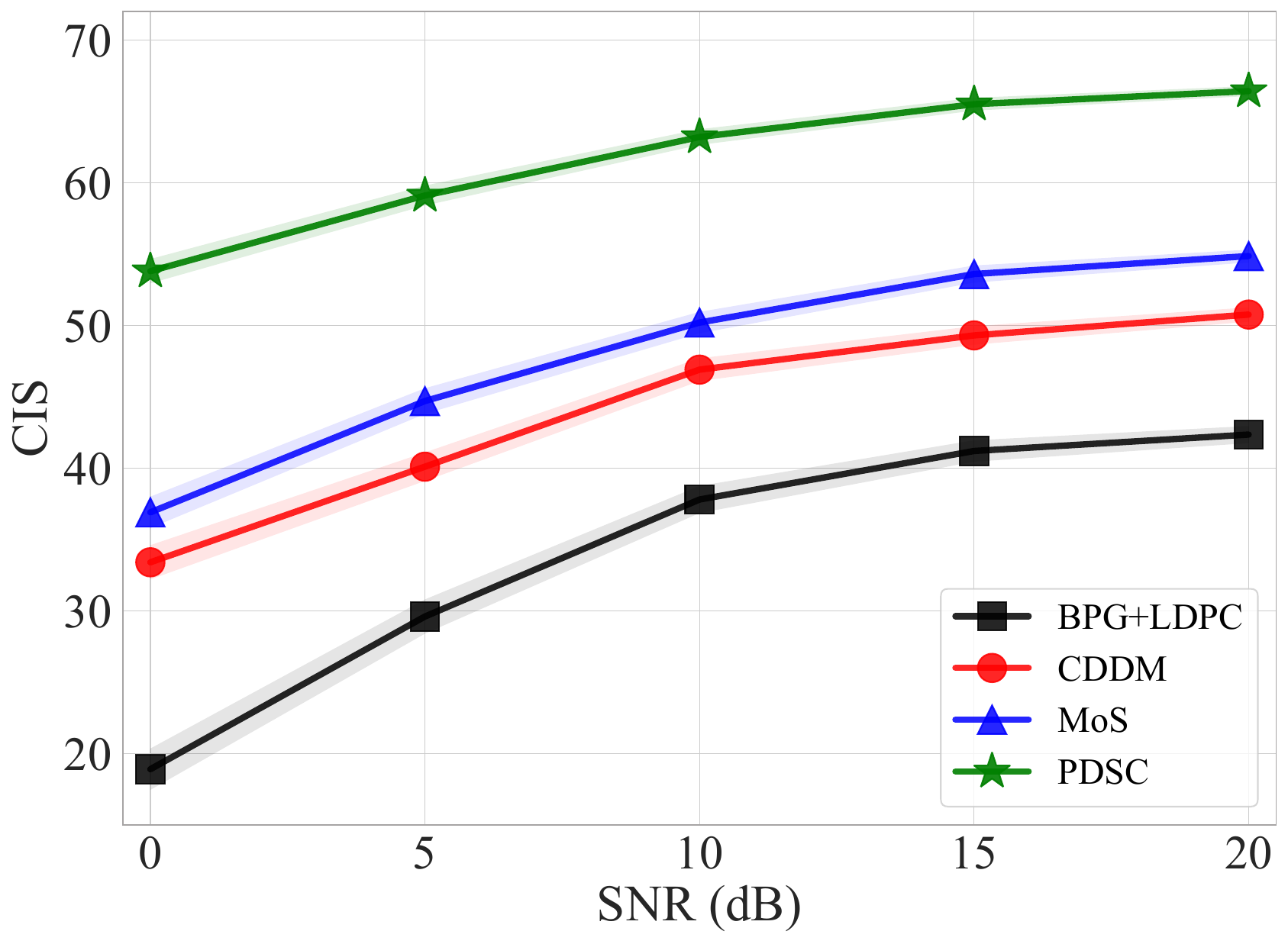}}
    \subfigure[CS versus SNR.]{
    \includegraphics[width=0.4\textwidth]{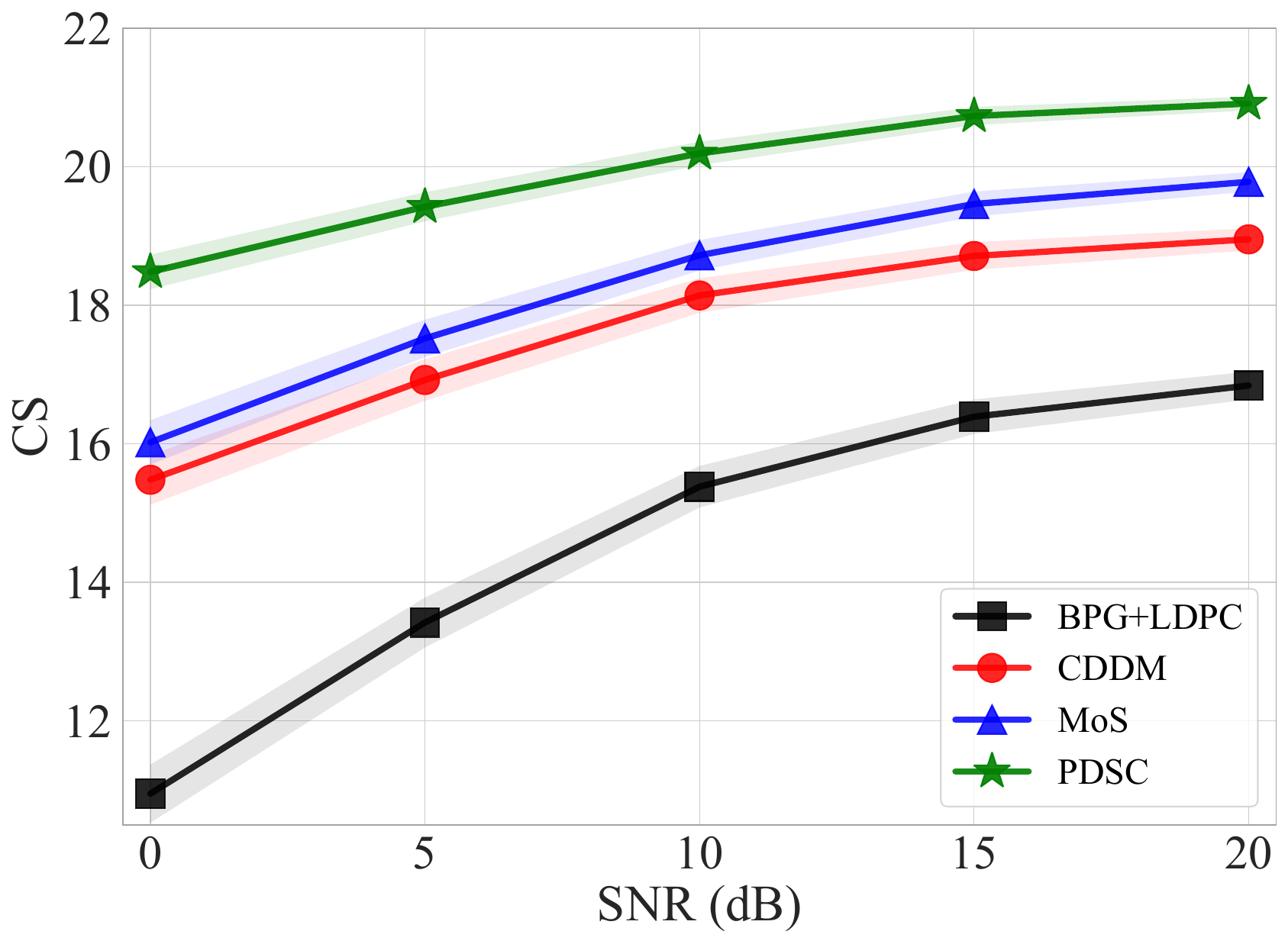}}
    \caption{Robustness comparison under different channel SNRs.}
    \label{fig:snr}
\end{figure}

\subsection{Robustness to Channel SNR}
Fig.~\ref{fig:snr} examines the robustness of different schemes under different channel SNRs, with the semantic source rate fixed at 210 bits/image. As expected, both CIS and CS increase with the SNR, since a more reliable channel improves semantic index recovery. PDSC consistently outperforms all baselines, especially in the low-SNR regime. Compared with the strongest baseline MoS, PDSC improves CIS by 45.8\% and CS by 15.4\% at 0 dB, and still achieves 21.1\% and 5.7\% gains at 20 dB, respectively. Averaged over all tested SNRs, PDSC provides 28.2\% higher CIS and 9.0\% higher CS than MoS. This improvement is attributed to the compact receiver-conditioned semantic indices, which are more robust to channel perturbations and more relevant to personalized reconstruction. In contrast, BPG+LDPC suffers from residual bit errors and error propagation, while CDDM and MoS do not explicitly model receiver preference. Moreover, CS varies more smoothly than CIS, indicating that personalized semantic alignment is more sensitive to the accurate recovery of preference-related information.

\begin{figure}[t]
    \centering
    \subfigure[CIS versus semantic rate.]{
    \includegraphics[width=0.4\textwidth]{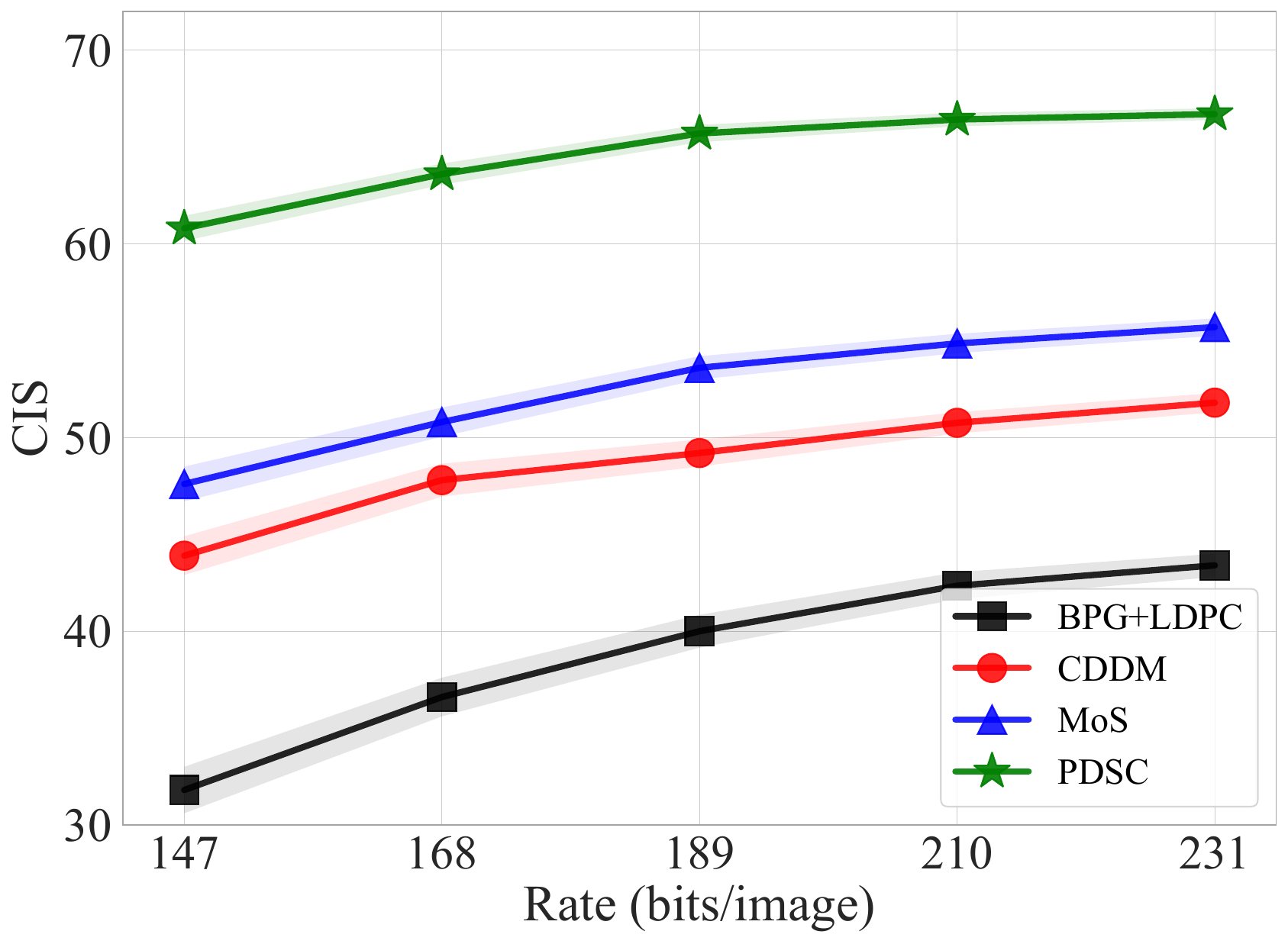}}
    \subfigure[CS versus semantic rate.]{
    \includegraphics[width=0.4\textwidth]{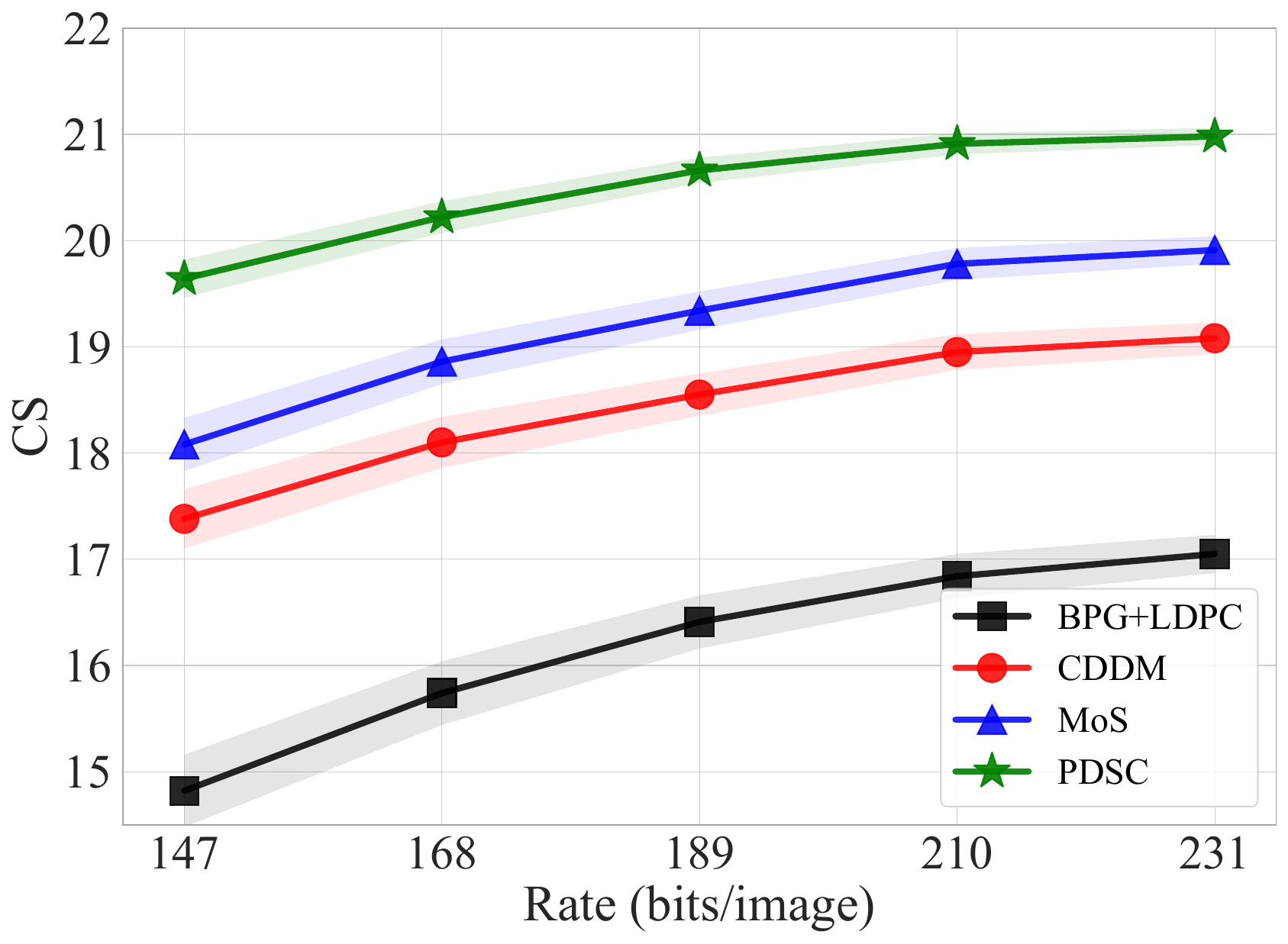}}
    \caption{Rate-distortion comparison vs. different semantic rates.}
    \label{fig:rate}
\end{figure}

\subsection{Rate-Distortion Analysis}
We next evaluate the rate-distortion performance under different semantic bit budgets. The SNR is fixed at 20 dB. For PDSC, $L$ is fixed to 21, while $K$ varies in $\{128,256,512,1024,2048\}$, corresponding to 147, 168, 189, 210, and 231 bits/image, respectively, according to
\begin{equation}
    R_s = L\lceil \log_2 K\rceil .
\end{equation}
All baselines are tested under the same bit budgets.

Fig.~\ref{fig:rate} presents the CIS and CS results versus the semantic source rate. It can be observed that both metrics improve as the rate increases, because a larger codebook reduces semantic quantization distortion and preserves richer source information. PDSC achieves the best performance at all considered rates, demonstrating its superior rate-distortion efficiency. Compared with MoS, PDSC improves CIS by 27.7\% and CS by 8.6\% at the lowest rate of 147 bits/image. At the default rate of 210 bits/image, PDSC still achieves 21.1\% higher CIS and 5.7\% higher CS than MoS. Averaged over all tested rates, PDSC provides 23.1\% and 6.7\% gains in CIS and CS, respectively. The reason is that PDSC allocates the limited bit budget to receiver-aware and task-relevant semantics, rather than transmitting redundant low-level details. Compared with BPG+LDPC, PDSC avoids the inefficiency of pixel-level compression for personalized reconstruction. Compared with CDDM and MoS, PDSC explicitly exploits receiver preference, thereby generating images that are more consistent with the personalized semantic objective. Furthermore, the performance gain tends to saturate at high semantic rates. In particular, increasing the codebook size beyond $K=1024$ provides only marginal improvement, indicating that most useful personalized semantics have already been captured.

\section{Conclusion}

In this paper, we proposed a PDSC framework for personalized wireless image transmission. Unlike existing user-agnostic semantic communication schemes, PDSC exploits users' historical interactions at the transmitter to extract source-aware personalized semantic tokens. To enable digital transmission, we designed a vector-quantized fixed-length semantic interface to convert continuous semantic tokens into a compact bitstream. At the receiver, a latent diffusion-based semantic decoder reconstructs personalized images from the recovered semantic tokens. We further introduced a personalized semantic distortion metric to jointly evaluate source-semantic fidelity and user-preference alignment. Experimental results demonstrated that PDSC achieves superior semantic quality and  personalization over existing baselines, particularly under limited bandwidth  conditions.

\vfill

\bibliographystyle{IEEEtran}
\bibliography{main}

\end{document}